\documentclass[aps,prl,twocolumn,groupedaddress,showpacs]{revtex4}
\usepackage{graphicx}
\usepackage{amsmath}
\usepackage{bm}

\begin{document}

\title{Competing mechanisms for singlet--triplet transition
in artificial molecules}

\author{Devis Bellucci}
\email[]{dbellucci@unimore.it}
\homepage[]{www.nanoscience.unimore.it}
\affiliation{INFM -  National Research Center on nano-Structures
and bio-Systems at Surfaces ($S^{3}$) and \\ Dipartimento di
Fisica, Universit\`a degli Studi di Modena e Reggio Emilia, Via
Campi 213/A, 41100 Modena, Italy}
\author{Massimo Rontani}
\affiliation{INFM - National Research Center on nano-Structures
and bio-Systems at Surfaces ($S^{3}$) and \\ Dipartimento di
Fisica, Universit\`a degli Studi di Modena e Reggio Emilia, Via
Campi 213/A, 41100 Modena, Italy}
\author{Filippo Troiani}
\affiliation{INFM - National Research Center on nano-Structures
and bio-Systems at Surfaces ($S^{3}$) and \\ Dipartimento di
Fisica, Universit\`a degli Studi di Modena e Reggio Emilia, Via
Campi 213/A, 41100 Modena, Italy}
\author{Guido Goldoni}
\affiliation{INFM - National Research Center on nano-Structures
and bio-Systems at Surfaces ($S^{3}$) and \\ Dipartimento di
Fisica, Universit\`a degli Studi di Modena e Reggio Emilia, Via
Campi 213/A, 41100 Modena, Italy}
\author{Elisa Molinari}
\affiliation{INFM - National Research Center on nano-Structures
and bio-Systems at Surfaces ($S^{3}$) and \\  Dipartimento di
Fisica, Universit\`a degli Studi di Modena e Reggio Emilia, Via
Campi 213/A, 41100 Modena, Italy}

\date{\today}

\begin{abstract}
We study the magnetic field induced singlet/triplet transition for
two electrons in vertically-coupled quantum dots by exact
diagonalization of the Coulomb interaction. We identify the
different mechanisms occurring in the transition, involving either
in-plane correlations or localization in opposite dots, depending
on the field direction. Therefore, both spin and orbital degrees
of freedom can be manipulated by field strength and direction. The
phase diagram of realistic devices is determined.
\end{abstract}


\maketitle

Atomic-like phenomenology, ensuing from the discrete density of
states, has been predicted and demonstrated in semiconductor
quantum dots (QDs), such as shell
structure~\cite{Ashoori96,Tarucha96}, fine structure due to
exchange interaction (Hund's rule)~\cite{Tarucha00}, and Kondo
physics~\cite{Goldhaber-Gordon98}; hence QDs are often termed
\emph{artificial atoms}. Carriers can be injected one by one into
the system in single-electron transport~\cite{Tarucha96} or
capacitance~\cite{Ashoori96} experiments, based on the {\em
Coulomb blockade}~\cite{Grabert92} phenomenon, and the energy
required to add one electron can be measured if the electrostatic
screening is poor and the thermal smearing is low.

Coupled QDs extend to the molecular realm the similarity between
natural and artificial atoms~\cite{Kouwenhoven95,Rontani01}; here,
inter-dot tunnelling introduces an energy scale which may be
comparable to other energy scales of the system, namely,
single-particle confinement energy, carrier-carrier interaction,
and magnetic energy. In contrast to natural molecules, where
inter-nuclear coupling is fixed by the balance between nuclear
repulsion and electrostatic attraction mediated by valence
electrons, in such \emph{artificial molecules} (AM) all energy
scales, including inter-dot coupling, as well as the charging
state of the system can be controlled to a very high degree by
device engineering and/or external fields~\cite{vanderWiel03}.

A typical AM consists of a disc-like region obtained from coupled
two-dimensional quantum systems, such as two quantum wells
(vertically coupled QDs). As in single QDs, electronic states can
be easily manipulated by a magnetic field $B_\perp$, perpendicular
to the plane of the QDs, which drives the system from a
low-correlation (low-field) regime to a highly correlated
(high-field) regime by changing the single-particle
splittings~\cite{Rontani02}. The study of electronic states of few
electrons in AMs~\cite{Palacios95,ReimannRMP} has become a topic
of increasing interest, partially due to possible implications for
the implementation of scalable solid-state quantum gates, with the
quantum bit of information coded either in the electron
charge~\cite{Barenco95} or spin~\cite{Loss98} degree of freedom
(DOF).

It should be noted that in AMs carriers are not only
electrostatically coupled, but also have their spin interlaced
when tunnelling is allowed~\cite{Burkard00}, since electrons with
opposite spin may tunnel into the same dot if the intra-dot
Coulomb interaction is not too large; the same process is
obviously prohibited for electrons with parallel spins. This
two-electron dynamics may be described by an effective Heisenberg
Hamiltonian $H = J({\bf B}) {\bf s}_1\cdot{\bf s}_2$ between spins
${\bf s}_1$ and ${\bf s}_2$~\cite{Burkard00}, with singlet and
triplet configurations separated by a field-dependent {\em
exchange-energy} gap $J ({\bf B})\equiv E_t -E_s$, which is
positive at zero field~\cite{exchange}. One convenient way to
control inter-dot tunnelling, and, hence, effective spin-spin
interaction $J$, is by applying a magnetic field with a finite
component in the plane of the QDs, i.e., perpendicular to the
tunnelling direction $B_\parallel$~\cite{Tokura00,Burkard00}.
Controlling tunnelling by $B_\parallel$ has the advantage that
other energy scales and, in particular, the Coulomb interaction
are practically unaffected. However, few studies are devoted to
this field configuration, which lacks the cylindrical symmetry
which can be exploited in the vertical field arrangement. On the
other hand, controlling $J({\bf B})$ in AM is crucial for the
proposed implementation of scalable quantum
gates~\cite{Burkard00}.

In this paper we study the exchange energy for two electrons
confined in AMs in a magnetic field of arbitrary direction. This
is performed by a fully numerical, real-space approach which
allows to account for the complexity of realistic samples; the
carrier-carrier Coulomb Hamiltonian is diagonalized exactly within
a large single-particle basis. We show that the field drives the
system from an uncorrelated regime, where the singlet state is
stable, to a strongly correlated one, where triplet ordering is
favored; however, the transition occurs by different mechanisms,
whether the field is in the vertical or in the in-plane direction.

We consider two electrons in a general QD structure. Carriers are
described by the effective-mass Hamiltonian
\begin{eqnarray}\label{eq:Hamiltonian}
    H & = & \sum^N_{i=1} \left[-\frac{\hbar^2}{2m^*}
    \left({\nabla}_i + \frac{e}{c}\mathbf{A}(\mathbf{r}_i)\right)^2
     + V(\mathbf{r}_i)\right] \nonumber\\
    & & + \frac{1}{2} \sum^N_{\substack{i,j=1 \\ i\neq j}}\frac{e^2}{\epsilon^*
    |\mathbf{r}_i-\mathbf{r}_j|} + g^*\mu_B {\bf B}\cdot{\bf S}
\end{eqnarray}
with $N = 2$. Here $m^*$, $\epsilon^*$, and $g^*$ are the
effective mass, dielectric constant, and $g-$factor,
respectively~\cite{parameters}. Equation (\ref{eq:Hamiltonian})
neglects non-parabolicity effects, but otherwise includes the full
3D nature of the quantum states in realistic samples, such as
layer width and finite band offsets, by the effective potential
$V(\textbf{r})$. Our numerical approach consists in mapping the
single-particle terms in a real-space grid, leading to a large
sparse matrix which is diagonalized by the Lanczos method.
Single-particle spin-orbitals are then used to build a basis of
Slater determinants for the $N$-particle problem, which is then
used to represent the two-body term, in the familiar Configuration
Interaction approach~\cite{Rontani01}. Coulomb matrix elements are
calculated numerically. The ensuing matrix, which can be very
large, is again sparse and can be diagonalized via the Lanczos
method as well~\cite{calcolo}.

In the following the potential $V(\mathbf{r})$ describes two
identical vertically coupled disk-like QDs. As usual for this type
of samples which have very different confinement energies in the
growth and in-plane directions, we separate the potential as
$V(x,y)+V(z)$, where $V(z)$ represent two symmetric quantum wells
of width $L_W$ separated by a barrier $L_B$ and conduction band
mismatch $V_0$. We perform the common choice of a parabolic
in-plane confinement $(1/2)m^*\omega^2_0 (x^2+y^2)$, as this has
proved to be quantitatively accurate~\cite{ReimannRMP}. Note,
however, that our numerical approach does not assume any symmetry;
in particular, the vector potential $\mathbf{A}(\mathbf{r})$ is
not limited to describe $z$-directed field.

\begin{figure}
  \includegraphics[width=9truecm]{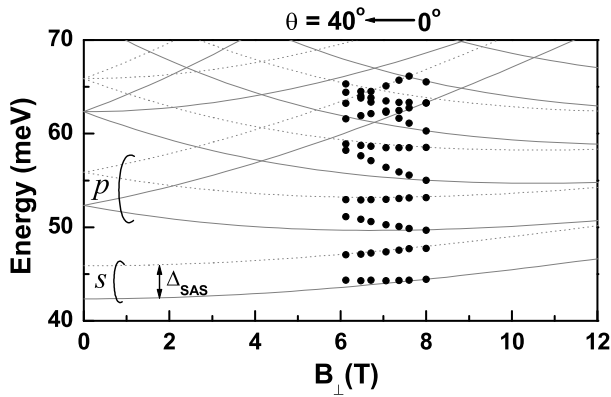}\\
  \caption{Single-particle energy levels for a GaAs AM in a magnetic field.
  Solid
  and dotted lines represent the FD states
  $\varepsilon_{nm}$ induced by a strictly vertical field
  for S and AS levels, respectively. $s$ and $p$ shells
  (see Ref.\ \onlinecite{molecular-notation}) are indicated.
  Dots represent calculated energy levels for a total field of 8 T,
  rotated from
  $0^\circ$ to $40^\circ$ with respect to the AM vertical axis.
  Sample parameters are as follows: $L_{W} =
  10\,\mbox{nm}$, $L_{B} = 3 \,\mbox{nm}$, $V_{0} = 300\,\mbox{meV}$, and
  $\hbar\omega_{0} = 10\,\mbox{meV}$.}
  \label{fig:b_perp}
\end{figure}

In a QD with parabolic in-plane confinement and strictly
perpendicular magnetic field, single-particle states are given by
the Fock-Darwin (FD) states (see, e.g.,
Ref.~\onlinecite{Merkt91}), with energies $ \varepsilon_{nm} =
\hbar\Omega (2n+|m|+1) -(\hbar\omega_c/2) m$,
 $n$ and $m$ being the principal and azimuthal quantum numbers,
respectively. The oscillator frequency is $\Omega =
\sqrt{\omega_0^2+\omega_c^2/4}$, with the cyclotron frequency
$\omega_c= eB/m^{*} c$. In symmetric AMs we have two such ladders
of energy levels, associated with the symmetric (S) and
anti-symmetric (AS) states arising from the double-well potential
in the growth direction, rigidly separated by a splitting
$\Delta_{\mbox{\scriptsize SAS}}$ (see Fig.\ \ref{fig:b_perp}).


We next consider the effect of a magnetic field with a finite
in-plane component $B_\parallel$. As shown in Fig.\
\ref{fig:b_perp}, when the angle $\theta$ between a fixed
$\left|{\bf B}\right|$ and the $z$ axis is increased, the energy
levels no longer correspond to the FD states at the corresponding
$B_\perp$. Indeed, the splitting between S and AS levels decreases
with increasing $\theta$~\cite{symmetry}, which shows that an
in-plane component of the field suppresses the tunnelling; note
that this effect is larger for higher levels. It is important to
stress that the in-plane field can meaningfully affect the motion
along the growth direction if $\omega_c^\parallel = e
B_\parallel/m^*c\sim\Delta_{\mbox{\scriptsize SAS}}$. Similar
effects are much harder to achieve in single QDs, due to the large
single-particle gaps induced by the single quantum well
confinement.


As discussed in more detail in the following, the reduction of the
energy gap between the $s$ and the $p$
shells~\cite{molecular-notation} (see Fig.\ \ref{fig:b_perp})
strongly reduces the single-particle energy of the triplet state
with respect to that of the singlet: the perpendicular field thus
promotes the singlet-triplet crossing. This transition results in
an enhancement of the in-plane correlation of the two-electron
groundstate and in the spin-polarization of the
system~\cite{Merkt91,Pfannkuche93}, arising from the exchange
(orbital) interaction~\cite{wigner}. Note that this mechanism only
involves the in-plane DOFs, and is therefore present in both
single and coupled QDs; in order to observe some marked
differences in the behavior of the two systems, one needs to
excite the motion along the growth direction $z$.

\begin{figure}
\includegraphics[width=9truecm]{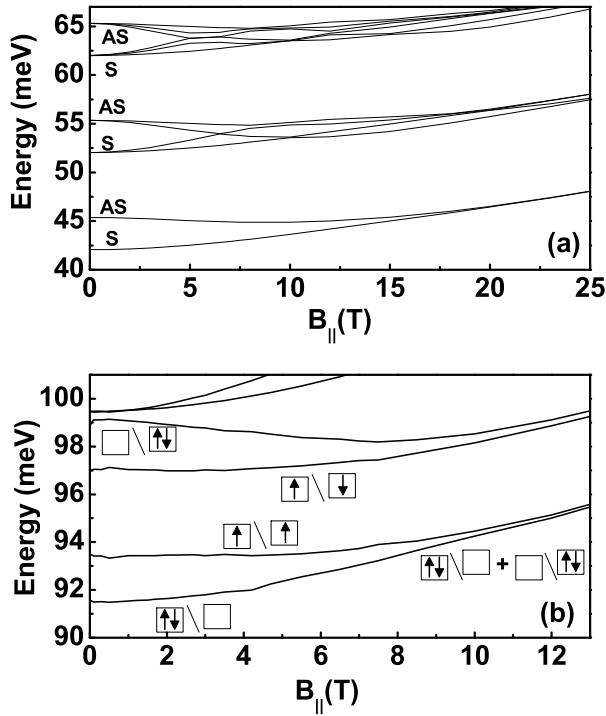}%
\caption{Energy levels vs in-plane field at $B_\perp = 0$ for the
same AM of Fig.\ \ref{fig:b_perp}. (a) Single-particle levels,
with indication of the S/AS character at low field. (b)
Two-electron levels. Insets: main components of the wavefunctions
in terms of S (left boxes) and AS (right boxes) single-particle
states.\label{fig:b_parallel}}
\end{figure}

Figure \ref{fig:b_parallel}(a) shows the single-particle levels as
a function of the in-plane field $B_\parallel$. The energy levels
come in shells with S and AS character, but the degeneracies which
are present at $B_\parallel=0$ are removed by a finite field, as
the axial symmetry of the system is lost. Therefore, the
single-particle wavefunctions do not have a well defined angular
momentum, and are now S or AS only with respect to a $180^\circ$
rotation about the axis parallel to ${\bf B}$. Besides, as the
field is increased, the S and AS levels approach each other, since
the tunnelling is progressively suppressed~\cite{Tokura00}.

In Fig.\ \ref{fig:b_parallel}(b) we show the lowest two-particle
levels, and schematically indicate the main components of the
corresponding wavefunctions in terms of S and AS single-particle
states. At low $B_\parallel$ the ground- and the first
excited-state have a singlet and a triplet character,
respectively. As $B_\parallel$ is increased, the energy gap $J$ is
suppressed: indeed, singlet and triplet states have the same
orbital energy, while the Zeeman term favors the latter (in the
field range of Fig.\ \ref{fig:b_parallel} the Zeeman contribution
can hardly be distinguished). As shown in the insets, the
transition occurs with the maximal, Coulomb-induced mixing of the
S and AS states, which is favored by the vanishing of
$\Delta_{\mbox{\scriptsize SAS}}$ at large fields. In other words,
increasing $B_\parallel$ the singlet state evolves from a nearly
pure S state to a fully entangled state in the S/AS basis. Note
also that, contrary to the one occurring at large $B_\perp$, here
the transition is associated to the correlation along the growth
direction, i.e., with the two electrons sitting on opposite QDs,
as we will show below.

\begin{figure}
  \includegraphics[width=95truemm]{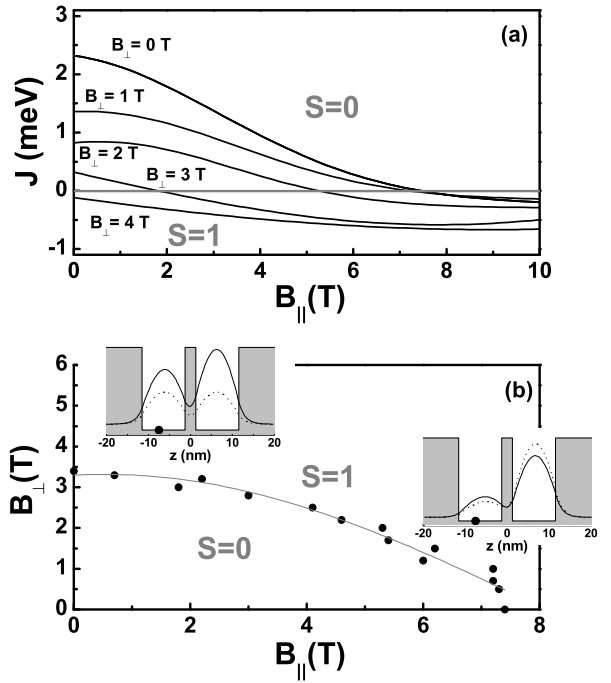}\\
  \caption{(a) Exchange energy $J$ vs in-plane field at selected vertical
  fields, for a GaAs AM. Sample parameters are as in Fig.\ \ref{fig:b_perp},
  but for a weaker lateral confinement $\hbar\omega_{0} = 4\,\mbox{meV}$.
  For clarity we show the best fitting curves from a large number
  of calculated points. Numerical inaccuracies may result in $\pm
  0.3$ meV shift from the curves only for the highest $B_\perp$.
  (b) Calculated singlet/triplet phase
  diagram. The line is a guide to the eye through the calculated points~\cite{inaccuracies}.
  Insets show the
  singlet (solid line) and triplet (dashed line) conditional
  probability near the transitions, defined as
    $ \left|\psi({\bm\rho}_0,z_0,{\bm\rho};z)\right|^2 $;
    $\psi$ is the two-electron wavefunction, with ${\bm\rho}$ the in-plane
  coordinate with respect to the vertical axis of the cylindrical QD.
  The reference electron (black dot) is fixed at $z_0 =
  -7.5\,\mbox{nm}$, at an in-plane position $|{\bm\rho}_0| = 4.4\,\mbox{nm}$;
  the conditional probability is then plotted
  along an axis parallel to $z$ and crossing the QD plane at a position diametrically
  opposed to the reference electron. Left inset: $B_\perp = 4\,\mbox{T}$,
  $B_\parallel = 0\,\mbox{T}$. Right inset: $B_\perp = 0\,\mbox{T}$,
  $B_\parallel = 9\,\mbox{T}$. }
  \label{fig:phase_diagram}
\end{figure}

The ability to control both the exchange energy $J$ and the
effective Hilbert-space structure is indeed pivotal to the
QD-based implementations of quantum-information
processing~\cite{Burkard00}; besides, $J$ is of direct
experimental interest, for it can be probed by single-electron
excitation spectroscopy~\cite{Leo97}. In Fig.\
\ref{fig:phase_diagram}(a) we show the calculated exchange energy
as a function of the in-plane field at different values of
$B_\perp$ and for a weaker parabolic confinement
($\Delta_{\scriptstyle SAS} \lesssim\hbar\omega_0$). The
positive/negative $J$ region is the stability region for
singlet/triplet states. Figure \ref{fig:phase_diagram}(a) shows
that an increase in $B_\perp$ monotonically $(i)$ reduces the
singlet stability range with respect to $B_\parallel$, and $(ii)$
enhances the ferromagnetic ($J<0$) behavior in the considered
range of $B_\parallel$ values. These features are summarized in
the phase diagram shown in Fig.\ \ref{fig:phase_diagram}(b).

A closer inspection into our results shows that different
mechanisms are involved in the singlet/triplet transition,
depending on the field direction. At zero field the singlet state
mainly corresponds to both electrons occupying the ($s$,S)
orbital, while a minor contribution from the ($s$,AS) orbital
gives rise to the spatial correlation in the $z$ direction. On the
contrary, all the dominant configurations in the triplet state
involve S states (see the conditional probability in Fig.\
\ref{fig:phase_diagram}(b)). $B_{\perp}$ leaves unaffected the $z$
DOF, while it energetically lowers the $p$ ($m=1$) state with
respect to the $s$ one. The positive single-particle contribution
to $J$ is therefore reduced, until it is compensated by the
negative contribution arising from the Coulomb energy. To
summarize, the singlet/triplet crossing induced by $B_\perp$ is
mainly connected with the in-plane dynamics, while it leaves
unaffected the motion in the growth direction and the
double-occupancy probability of each dot.

The main effect of $B_\parallel$ (right inset), instead, is that
of suppressing the energy splitting resulting from the interdot
tunnelling. This clearly favors the occupation of the AS states,
and therefore vertical correlations for both the singlet and the
triplet states set in; in both cases, the two electrons tend to
localize in opposite dots, and the importance of the spins'
relative orientation vanishes with the double occupancy
probability. Indeed the exponential vanishing of $J$ represents
the clear fingerprint of the regime where the double-occupancy
probability is suppressed. It should be noted that this is not a
single-particle effect, since it does not imply, nor require, the
complete suppression of the tunnelling.

The results reported in Fig.\ \ref{fig:phase_diagram} show that
these two different mechanisms interfere with each other in a
non-trivial manner. The presence of the perpendicular component
$B_\perp$ favors the single-triplet crossing and the ferromagnetic
phase, while it opposes the suppression of the double occupancy
and the resulting singlet-triplet degeneracy (apart from the
Zeeman term). Such interplay arises from the 3D nature of quantum
states in the AM: in fact, in the considered range of physical
parameters ($\hbar\omega_0\sim\Delta_{\mbox{\scriptsize SAS}}$),
the magnetic field can strongly affect both the in-plane
(intra-dot) and the vertical (inter-dot) DOF.

An adiabatic manipulation of $J$ by means of magnetic (and
electric) fields has been proposed in order to implement the
two-qubit gates in electron-spin based quantum
computers~\cite{Burkard00}. The rest condition within such scheme
would correspond to the suppression of $J$ \emph{and} of the
overlap between electrons localized in adjacent QDs, where both
conditions should be induced by a static magnetic field. In this
perspective, our findings suggest that $(i)$ the $B_{\parallel}$
(rather than $B_{\perp}$) component of the field and the
exponential suppression (rather than the crossing point from $J>0$
to $J<0$) are required; $(ii)$ the presence of a field component
perpendicular to the static one (as required, e.g., for the
single-spin rotations) should be simultaneously taken into account
in order to determine the suited range of physical parameters.

To summarize, we have theoretically investigated the dependence of
the singlet and triplet states of two electrons in AMs on external
magnetic fields of arbitrary direction. Our computational approach
allows to fully account for the different physical mechanisms
underlying the singlet/triplet transitions which are due to the
parallel and perpendicular components of the field, as well as for
the non-trivial interplay between the vertical and the in-plane
correlation effects that they induce. The perpendicular component
of the field does indeed facilitate the transition from the
anti-ferromagnetic to the ferromagnetic phase which is induced by
the parallel component, but at the same time it opposes the
carrier localization and correlation properties that the latter
tends to induce in the AM.


We are grateful to A. Bertoni, U.~Hohenester, V.~Pellegrini, and
C.~Tejedor for useful discussions. This work was supported in part
by MIUR FIRB-RBAU01ZEML \textsl{Quantum phases of ultra-low
electron density semiconductor heterostructures}, and by INFM-IT
\textsl{Calcolo Parallelo} (2003). With the contribution of the
Italian Minister for Foreign Affairs, Direzione Generale per la
Promozione e la Cooperazione Culturale.



\begin{thebibliography}{32}
\expandafter\ifx\csname
natexlab\endcsname\relax\def\natexlab#1{#1}\fi
\expandafter\ifx\csname bibnamefont\endcsname\relax
  \def\bibnamefont#1{#1}\fi
\expandafter\ifx\csname bibfnamefont\endcsname\relax
  \def\bibfnamefont#1{#1}\fi
\expandafter\ifx\csname citenamefont\endcsname\relax
  \def\citenamefont#1{#1}\fi
\expandafter\ifx\csname url\endcsname\relax
  \def\url#1{\texttt{#1}}\fi
\expandafter\ifx\csname
urlprefix\endcsname\relax\def\urlprefix{URL }\fi
\providecommand{\bibinfo}[2]{#2}
\providecommand{\eprint}[2][]{\url{#2}}

\bibitem[{\citenamefont{Ashoori}(1996)}]{Ashoori96}
\bibinfo{author}{\bibfnamefont{R.}~\bibnamefont{Ashoori}},
  \bibinfo{journal}{Nature} \textbf{\bibinfo{volume}{379}},
  \bibinfo{pages}{413} (\bibinfo{year}{1996}).

\bibitem[{\citenamefont{Tarucha et~al.}(1996)\citenamefont{Tarucha, Austing,
  Honda, van~der Hage, and Kouwenhoven}}]{Tarucha96}
\bibinfo{author}{\bibfnamefont{S.}~\bibnamefont{Tarucha}},
  \bibinfo{author}{\bibfnamefont{D.~G.} \bibnamefont{Austing}},
  \bibinfo{author}{\bibfnamefont{T.}~\bibnamefont{Honda}},
  \bibinfo{author}{\bibfnamefont{R.~J.} \bibnamefont{van~der Hage}},
  \bibnamefont{and} \bibinfo{author}{\bibfnamefont{L.~P.}
  \bibnamefont{Kouwenhoven}},
\bibinfo{journal}{Phys.\ Rev.\ Lett.}
  \textbf{\bibinfo{volume}{77}}, \bibinfo{pages}{3613} (\bibinfo{year}{1996}).

\bibitem[{\citenamefont{Tarucha et~al.}(2000)\citenamefont{Tarucha, Austing,
  Tokura, van~der Wiel, and Kouwenhoven}}]{Tarucha00}
\bibinfo{author}{\bibfnamefont{S.}~\bibnamefont{Tarucha}},
  \bibinfo{author}{\bibfnamefont{D.~G.} \bibnamefont{Austing}},
  \bibinfo{author}{\bibfnamefont{Y.}~\bibnamefont{Tokura}},
  \bibinfo{author}{\bibfnamefont{W.~G.} \bibnamefont{van~der Wiel}},
  \bibnamefont{and} \bibinfo{author}{\bibfnamefont{L.~P.}
  \bibnamefont{Kouwenhoven}},
\bibinfo{journal}{Phys.\ Rev.\ Lett.}
  \textbf{\bibinfo{volume}{84}}, \bibinfo{pages}{2485} (\bibinfo{year}{2000}).

\bibitem[{\citenamefont{Goldhaber-Gordon
  et~al.}(1998)\citenamefont{Goldhaber-Gordon, Shtrikman, Mahalu,
  Abusch-Magder, Meirav, and Kastner}}]{Goldhaber-Gordon98}
\bibinfo{author}{\bibfnamefont{D.~G.} \bibnamefont{Goldhaber-Gordon}},
  \bibinfo{author}{\bibfnamefont{H.}~\bibnamefont{Shtrikman}},
  \bibinfo{author}{\bibfnamefont{D.}~\bibnamefont{Mahalu}},
  \bibinfo{author}{\bibfnamefont{D.}~\bibnamefont{Abusch-Magder}},
  \bibinfo{author}{\bibfnamefont{U.}~\bibnamefont{Meirav}}, \bibnamefont{and}
  \bibinfo{author}{\bibfnamefont{M.~A.} \bibnamefont{Kastner}},
  \bibinfo{journal}{Nature} \textbf{\bibinfo{volume}{391}},
  \bibinfo{pages}{156} (\bibinfo{year}{1998});
\bibinfo{author}{\bibfnamefont{S.~M.} \bibnamefont{Cronenwett}},
  \bibinfo{author}{\bibfnamefont{T.~H.} \bibnamefont{Oosterkamp}},
  \bibnamefont{and} \bibinfo{author}{\bibfnamefont{L.~P.}
  \bibnamefont{Kouwenhoven}}, \bibinfo{journal}{Science}
  \textbf{\bibinfo{volume}{281}}, \bibinfo{pages}{540} (\bibinfo{year}{1998}).

\bibitem[{\citenamefont{Grabert and Devoret}(1992)}]{Grabert92}
  \emph{\bibinfo{title}{Single charge tunneling: Coulomb blockade phenomena in
  nanostructures}},  Edited by H. Grabert, M.H. Devoret
   (\bibinfo{publisher}{Plenum}, \bibinfo{address}{New York},
  \bibinfo{year}{1992}), \emph{\bibinfo{series}{NATO ASI series B}},
  Vol. \bibinfo{volume}{294}.

\bibitem[{\citenamefont{Kouwenhoven}(1995)}]{Kouwenhoven95}
\bibinfo{author}{\bibfnamefont{L.~P.} \bibnamefont{Kouwenhoven}},
  \bibinfo{journal}{Science} \textbf{\bibinfo{volume}{268}},
  \bibinfo{pages}{1440} (\bibinfo{year}{1995}).

\bibitem[{\citenamefont{Rontani et~al.}(2001)\citenamefont{Rontani, Troiani,
  Hohenester, and Molinari}}]{Rontani01}
\bibinfo{author}{\bibfnamefont{M.}~\bibnamefont{Rontani}},
  \bibinfo{author}{\bibfnamefont{F.}~\bibnamefont{Troiani}},
  \bibinfo{author}{\bibfnamefont{U.}~\bibnamefont{Hohenester}},
  \bibnamefont{and} \bibinfo{author}{\bibfnamefont{E.}~\bibnamefont{Molinari}},
  \bibinfo{journal}{Solid State Commun.} \textbf{\bibinfo{volume}{119}},
  \bibinfo{pages}{309} (\bibinfo{year}{2001})
.

\bibitem[{\citenamefont{van~del Wiel et~al.}(2003)\citenamefont{van~del Wiel,
  De~Franceschi, Elzerman, Fujisawa, Tarucha, and Kouwenhoven}}]{vanderWiel03}
\bibinfo{author}{\bibfnamefont{W.~G.} \bibnamefont{van~del Wiel}},
  \bibinfo{author}{\bibfnamefont{S.}~\bibnamefont{De~Franceschi}},
  \bibinfo{author}{\bibfnamefont{J.~M.} \bibnamefont{Elzerman}},
  \bibinfo{author}{\bibfnamefont{T.}~\bibnamefont{Fujisawa}},
  \bibinfo{author}{\bibfnamefont{S.}~\bibnamefont{Tarucha}}, \bibnamefont{and}
  \bibinfo{author}{\bibfnamefont{L.~P.} \bibnamefont{Kouwenhoven}},
  \bibinfo{journal}{Rev. Mod. Phys.} \textbf{\bibinfo{volume}{75}},
  \bibinfo{pages}{1} (\bibinfo{year}{2003}).

\bibitem[{\citenamefont{Rontani et~al.}(2002)\citenamefont{Rontani, Goldoni,
  Manghi, and Molinari}}]{Rontani02}
\bibinfo{author}{\bibfnamefont{M.}~\bibnamefont{Rontani}},
  \bibinfo{author}{\bibfnamefont{G.}~\bibnamefont{Goldoni}},
  \bibinfo{author}{\bibfnamefont{F.}~\bibnamefont{Manghi}}, \bibnamefont{and}
  \bibinfo{author}{\bibfnamefont{E.}~\bibnamefont{Molinari}},
  \bibinfo{journal}{Europhys.\ Lett.} \textbf{\bibinfo{volume}{58}},
  \bibinfo{pages}{555} (\bibinfo{year}{2002}).

\bibitem[{\citenamefont{Palacios and Hawrylak}(1995)}]{Palacios95}
\bibinfo{author}{\bibfnamefont{J.~J.} \bibnamefont{Palacios}} \bibnamefont{and}
  \bibinfo{author}{\bibfnamefont{P.}~\bibnamefont{Hawrylak}},
  \bibinfo{journal}{Phys.\ Rev.\ B} \textbf{\bibinfo{volume}{51}},
  \bibinfo{pages}{1769} (\bibinfo{year}{1995});
\bibinfo{author}{\bibfnamefont{J.~H.} \bibnamefont{Oh}},
  \bibinfo{author}{\bibfnamefont{K.~J.} \bibnamefont{Chang}},
  \bibinfo{author}{\bibfnamefont{G.}~\bibnamefont{Ihm}}, \bibnamefont{and}
  \bibinfo{author}{\bibfnamefont{S.~J.} \bibnamefont{Lee}},
  \bibinfo{journal}{Phys.\ Rev.\ B} \textbf{\bibinfo{volume}{53}},
  \bibinfo{pages}{R13264} (\bibinfo{year}{1996});
\bibinfo{author}{\bibfnamefont{M.}~\bibnamefont{Rontani}},
  \bibinfo{author}{\bibfnamefont{F.}~\bibnamefont{Rossi}},
  \bibinfo{author}{\bibfnamefont{F.}~\bibnamefont{Manghi}}, \bibnamefont{and}
  \bibinfo{author}{\bibfnamefont{E.}~\bibnamefont{Molinari}},
  \bibinfo{journal}{Solid State Commun.} \textbf{\bibinfo{volume}{112}},
  \bibinfo{pages}{151} (\bibinfo{year}{1999});
\bibinfo{author}{\bibfnamefont{Y.}~\bibnamefont{Tokura}},
  \bibinfo{author}{\bibfnamefont{D.~G.} \bibnamefont{Austing}},
  \bibnamefont{and} \bibinfo{author}{\bibfnamefont{S.}~\bibnamefont{Tarucha}},
  \bibinfo{journal}{J.\ Phys.: Condens.~Matter} \textbf{\bibinfo{volume}{11}},
  \bibinfo{pages}{6023} (\bibinfo{year}{1999});
\bibinfo{author}{\bibfnamefont{B.}~\bibnamefont{Partoens}} \bibnamefont{and}
  \bibinfo{author}{\bibfnamefont{F.~M.} \bibnamefont{Peeters}},
  \bibinfo{journal}{Phys.\ Rev.\ Lett.} \textbf{\bibinfo{volume}{84}},
  \bibinfo{pages}{4433} (\bibinfo{year}{2000});
\bibinfo{author}{\bibfnamefont{L.}~\bibnamefont{Martin-Moreno}},
  \bibinfo{author}{\bibfnamefont{L.}~\bibnamefont{Brey}}, \bibnamefont{and}
  \bibinfo{author}{\bibfnamefont{C.}~\bibnamefont{Tejedor}},
  \bibinfo{journal}{Phys.\ Rev.\ B} \textbf{\bibinfo{volume}{62}},
  \bibinfo{pages}{R10633} (\bibinfo{year}{2000});
\bibinfo{author}{\bibfnamefont{M.}~\bibnamefont{Pi}},
  \bibinfo{author}{\bibfnamefont{A.}~\bibnamefont{Emperador}},
  \bibinfo{author}{\bibfnamefont{M.}~\bibnamefont{Barranco}},
  \bibinfo{author}{\bibfnamefont{F.}~\bibnamefont{Garcias}},
  \bibinfo{author}{\bibfnamefont{K.}~\bibnamefont{Muraki}},
  \bibinfo{author}{\bibfnamefont{S.}~\bibnamefont{Tarucha}}, \bibnamefont{and}
  \bibinfo{author}{\bibfnamefont{D.~G.} \bibnamefont{Austing}},
  \bibinfo{journal}{Phys.\ Rev.\ Lett.} \textbf{\bibinfo{volume}{87}},
  \bibinfo{pages}{066801} (\bibinfo{year}{2001});
\bibinfo{author}{\bibfnamefont{M.}~\bibnamefont{Rontani}},
  \bibinfo{author}{\bibfnamefont{G.}~\bibnamefont{Goldoni}}, \bibnamefont{and}
  \bibinfo{author}{\bibfnamefont{E.}~\bibnamefont{Molinari}}, in
  \emph{\bibinfo{title}{New directions in mesoscopic physics (towards
  nanoscience)}}, edited by R. Fazio, V. F. Gantmakher, and Y. Imry
  (\bibinfo{publisher}{Kluwer}, \bibinfo{address}{Dordrecht},
  \bibinfo{year}{2003}), NATO ASI series B, Vol.~\bibinfo{volume}{125}, \bibinfo{pages}{361};
  \bibinfo{note}{cond-mat/0212626};
\bibitem[{\citenamefont{Reimann and Manninen}(2002)}]{ReimannRMP}
\bibinfo{author}{\bibfnamefont{S.~M.} \bibnamefont{Reimann}} \bibnamefont{and}
  \bibinfo{author}{\bibfnamefont{M.}~\bibnamefont{Manninen}},
  \bibinfo{journal}{Rev.\ Mod.\ Phys.} \textbf{\bibinfo{volume}{74}},
  \bibinfo{pages}{1283} (\bibinfo{year}{2002}).

\bibitem[{\citenamefont{Barenco et~al.}(1995)\citenamefont{Barenco, Deutsch,
  Ekert, and Jozsa}}]{Barenco95}
\bibinfo{author}{\bibfnamefont{A.}~\bibnamefont{Barenco}},
  \bibinfo{author}{\bibfnamefont{D.}~\bibnamefont{Deutsch}},
  \bibinfo{author}{\bibfnamefont{A.}~\bibnamefont{Ekert}}, \bibnamefont{and}
  \bibinfo{author}{\bibfnamefont{R.}~\bibnamefont{Jozsa}},
  \bibinfo{journal}{Phys.\ Rev.\ Lett.} \textbf{\bibinfo{volume}{74}},
  \bibinfo{pages}{4083} (\bibinfo{year}{1995}).

\bibitem[{\citenamefont{Loss and DiVincenzo}(1998)}]{Loss98}
\bibinfo{author}{\bibfnamefont{D.}~\bibnamefont{Loss}} \bibnamefont{and}
  \bibinfo{author}{\bibfnamefont{D.~P.} \bibnamefont{DiVincenzo}},
  \bibinfo{journal}{Phys. Rev. A} \textbf{\bibinfo{volume}{57}},
  \bibinfo{pages}{120} (\bibinfo{year}{1998}).

\bibitem[{\citenamefont{Burkard et~al.}(2000)\citenamefont{Burkard, Seelig, and
  Loss}}]{Burkard00}
\bibinfo{author}{\bibfnamefont{G.}~\bibnamefont{Burkard}},
  \bibinfo{author}{\bibfnamefont{G.}~\bibnamefont{Seelig}}, \bibnamefont{and}
  \bibinfo{author}{\bibfnamefont{D.}~\bibnamefont{Loss}},
  \bibinfo{journal}{Phys.\ Rev.\ B} \textbf{\bibinfo{volume}{62}},
  \bibinfo{pages}{2581} (\bibinfo{year}{2000}).

\bibitem[{cal()}]{exchange}
\bibinfo{note}{The {\em exchange energy} $J(B_\perp,B_\parallel)$ results from both
single-particle and Coulomb contributions. The term {\em exchange
interaction}, as used in the paper, indicates only the Coulomb
term depending on the symmetry of the two-electron spatial
wavefunction.}

\bibitem[{\citenamefont{Tokura et~al.}(2000)\citenamefont{Tokura, Sasaki,
  Austing, and Tarucha}}]{Tokura00}
\bibinfo{author}{\bibfnamefont{Y.}~\bibnamefont{Tokura}},
  \bibinfo{author}{\bibfnamefont{S.}~\bibnamefont{Sasaki}},
  \bibinfo{author}{\bibfnamefont{D.~G.} \bibnamefont{Austing}},
  \bibnamefont{and} \bibinfo{author}{\bibfnamefont{S.}~\bibnamefont{Tarucha}},
  \bibinfo{journal}{Physica E} \textbf{\bibinfo{volume}{6}},
  \bibinfo{pages}{676} (\bibinfo{year}{2000}).

\bibitem[{par()}]{parameters}
\bibinfo{note}{We use standard values of GaAs bulk parameters: $m^*/m_0 = 0.067$,
  $\epsilon^{*} = 12.9$, $g^* = -0.44$.}

\bibitem[{cal()}]{calcolo}
\bibinfo{note}{Tipical calculations are performed on a ($100\times 100\times
  40$) nm grid with ($64\times 64\times 128$) points along directions
  ($x$,$y$,$z$), respectively; the Coulomb interaction is represented on a
  single-particle basis set consisting of 12 states.}


\bibitem[{\citenamefont{Merkt et~al.}(1991)\citenamefont{Merkt, Huser, and
  Wagner}}]{Merkt91}
\bibinfo{author}{\bibfnamefont{U.}~\bibnamefont{Merkt}},
  \bibinfo{author}{\bibfnamefont{J.}~\bibnamefont{Huser}}, \bibnamefont{and}
  \bibinfo{author}{\bibfnamefont{M.}~\bibnamefont{Wagner}},
  \bibinfo{journal}{Phys.\ Rev.\ B} \textbf{\bibinfo{volume}{43}},
  \bibinfo{pages}{7320} (\bibinfo{year}{1991}).

\bibitem[{sym()}]{symmetry}
\bibinfo{note}{Here the S/AS labelling is used for brevity. Obviously, for a
  general field direction with respect to the tunnelling
  direction,
  wavefunctions do not have a well defined S/AS symmetry. See
  \textcite{Bellucci03physe}.}

\bibitem{molecular-notation} We classify the AM states
according to their vertical (S/AS) and in-plane (FD) character.
The $s$ and $p$ shells correspond to the $(n,m)=(0,0)$ and $(0,\pm
1)$ FD states, respectively. In the literature, AM states are
sometimes classified according to the molecular notation: the $s$
and $p$ states with S (AS) character correspond to $\sigma_g$,
$\pi_u$ ($\sigma_u$, $\pi_g$) molecular orbitals, respectively.
See \textcite{Rontani01}.

\bibitem[{\citenamefont{Pfannkuche et~al.}(1993)\citenamefont{Pfannkuche,
  Gudmundsson, and Maksym}}]{Pfannkuche93}
\bibinfo{author}{\bibfnamefont{D.}~\bibnamefont{Pfannkuche}},
  \bibinfo{author}{\bibfnamefont{V.}~\bibnamefont{Gudmundsson}},
  \bibnamefont{and} \bibinfo{author}{\bibfnamefont{P.~A.}
  \bibnamefont{Maksym}}, \bibinfo{journal}{Phys.\ Rev.\ B}
  \textbf{\bibinfo{volume}{47}}, \bibinfo{pages}{2244} (\bibinfo{year}{1993}).

\bibitem[{wig()}]{wigner}
\bibinfo{note}{Eventually, as the field is further increased, electron localize
  in a confined counterpart of a Wigner crystal. See \textcite{Rontani02}.}

\bibitem[{\citenamefont{Kouwenhoven et~al.}(1997)\citenamefont{Kouwenhoven,
  Oosterkamp, Danoesastro, Eto, Austing, Honda, and Tarucha}}]{Leo97}
\bibinfo{author}{\bibfnamefont{L.~P.} \bibnamefont{Kouwenhoven}},
  \bibinfo{author}{\bibfnamefont{T.~H.} \bibnamefont{Oosterkamp}},
  \bibinfo{author}{\bibfnamefont{M.~W.~S.} \bibnamefont{Danoesastro}},
  \bibinfo{author}{\bibfnamefont{M.}~\bibnamefont{Eto}},
  \bibinfo{author}{\bibfnamefont{D.~G.} \bibnamefont{Austing}},
  \bibinfo{author}{\bibfnamefont{T.}~\bibnamefont{Honda}}, \bibnamefont{and}
  \bibinfo{author}{\bibfnamefont{S.}~\bibnamefont{Tarucha}},
  \bibinfo{journal}{Science} \textbf{\bibinfo{volume}{278}},
  \bibinfo{pages}{1788} (\bibinfo{year}{1997}).

\bibitem[{\citenamefont{Bellucci et~al.}(2004)\citenamefont{Bellucci,
  Rontani, Goldoni, Troiani, and Molinari}}]{Bellucci03physe}
\bibinfo{author}{\bibfnamefont{D.}~\bibnamefont{Bellucci}},
  \bibinfo{author}{\bibfnamefont{M.}~\bibnamefont{Rontani}},
  \bibinfo{author}{\bibfnamefont{G.}~\bibnamefont{Goldoni}},
  \bibinfo{author}{\bibfnamefont{F.}~\bibnamefont{Troiani}}, \bibnamefont{and}
  \bibinfo{author}{\bibfnamefont{E.}~\bibnamefont{Molinari}},
  \bibinfo{journal}{Physica E} \textbf{\bibinfo{volume}{22}}
  \bibinfo{pages}{482}, (\bibinfo{year}{2004}).

\bibitem{inaccuracies} Scattering of calculated points is related
to the finiteness of the space grids. These inaccuracies are more
significant for non-symmetric field directions.

\end{thebibliography}

\end{document}